\documentclass{article}

\usepackage{PRIMEarxiv}

\usepackage[utf8]{inputenc} 
\usepackage[T1]{fontenc}    
\usepackage{hyperref}       
\usepackage{url}            
\usepackage{booktabs}       
\usepackage{amsfonts}       
\usepackage{nicefrac}       
\usepackage{microtype}      
\usepackage{lipsum}
\usepackage{natbib}
\usepackage{fancyhdr}       
\usepackage{graphicx}       
\graphicspath{{media/}}     
\title{Privacy Challenges and Solutions in RAG-Enhanced LLMs for Healthcare Chatbots: A Review of Applications, Risks, and Future Directions}

\author{
  Shaowei Guan, Hin Chi Kwok \\
  Centre for Smart Health, School of Nursing \\ 
  The Hong Kong Polytechnic University\\
  Hong Kong, China \\
   \And
  Ngai Fong Law \\
  Department of Electrical and Electronic Engineering\\
  The Hong Kong Polytechnic University\\
  Hong Kong, China\\
  \texttt{email@email} \\
   \AND
  Gregor Stiglic \\
  Faculty of Health Sciences\\
  University of Maribor\\
  Maribor, Slovenia\\
  \And
  Harry QIN, Vivian Hui \\
  Centre for Smart Health, School of Nursing \\ 
  The Hong Kong Polytechnic University\\
  Hong Kong, China \\
}

\begin{document}
\maketitle

\begin{abstract}
Retrieval-augmented generation (RAG) has rapidly emerged as a transformative approach for integrating large language models into clinical and biomedical workflows. However, privacy risks, such as protected health information (PHI) exposure, remain inconsistently mitigated. This review provides a thorough analysis of the current landscape of RAG applications in healthcare, including (i) sensitive data type across clinical scenarios, (ii) the associated privacy risks, (iii) current and emerging data-privacy protection mechanisms and (iv) future direction for patient data privacy protection. We synthesize 23 articles on RAG applications in healthcare and systematically analyze privacy challenges through a pipeline-structured framework encompassing data storage, transmission, retrieval and generation stages, delineating potential failure modes, their underlying causes in threat models and system mechanisms, and their practical implications. Building on this analysis, we critically review 17 articles on privacy-preserving strategies for RAG systems. Our evaluation reveals critical gaps, including insufficient clinical validation, absence of standardized evaluation frameworks, and lack of automated assessment tools. We propose actionable directions based on these limitations and conclude with a call to action. This review provides researchers and practitioners with a structured framework for understanding privacy vulnerabilities in healthcare RAG and offers a roadmap toward developing systems that achieve both clinical effectiveness and robust privacy preservation.
\end{abstract}

\keywords{Data Privacy \and Retrieval-Augmented Generation \and Large Language Models \and Privacy-Preserving \and Privacy Risks \and Healthcare Informatics \and Clinical Applications}
\section{Introduction}
Large Language Models (LLMs) are rapidly transforming healthcare delivery, supporting diverse applications from patient care and education to clinical documentation and decision support \cite{ho2025development}. However, their deployment in high-stakes clinical environments is constrained by well-documented limitations, such as hallucinations, inherent data biases, and the lack of explainability, which compromise safe and accountable use of LLMs. Retrieval-Augmented Generation (RAG) has emerged as a promising solution to this challenge by conditioning generation on curated information from authoritative sources, including clinical guidelines, textbooks, and medical records. RAG mitigates LLM limitations through a three-step process: (i) embedding and indexing vetted medical knowledge, (ii) retrieving context relevant to user queries, and (iii) conditioning the generator on that context \cite{mahapatra2025storage}. Early deployments in medical education, chronic disease counselling, and workflow assistance demonstrate substantial improvements in answer reliability and user trust compared with purely parametric LLMs \cite{xu2025development, kelly2025effectiveness, son2025development}.

Despite these advantages, the integration of RAG into healthcare workflows introduces unique and complex privacy challenges that extend beyond those encountered in traditional health information technology or standalone LLMs \cite{neha2025retrieval}. Healthcare RAG deployments are being integrated into high-stakes workflows, including triage chatbots, discharge counselling, guideline lookup, radiology consultation, and rare disease support, that routinely process patient-linked text, institution-specific documents, and population-level datasets \cite{busch2025evaluation, kreimeyer2024using}. Even when data repositories are considered "de-identified," they remain vulnerable to privacy breaches. Techniques such as embedding inversion attacks, prompt injection, and membership inference can re-identify individuals or re-expose sensitive information \cite{huang2024transferable, anderson2024my}. Consequently, privacy risk in healthcare RAG is not merely a narrow information technology concern but rather a system-wide property that directly affects clinical reliability, clinician adoption, and patient willingness to disclose information.

The incorporation of retrieval mechanisms into LLM pipelines fundamentally reshapes the privacy risk surface. Healthcare RAG systems may handle sensitive health information (SHI), including protected health information (PHI) and institution-specific knowledge (i.e., internal clinical protocols, supplier contracts, operational staffing models and metrics), across multiple architectural components (indexers, vector stores, re-rankers, generators) and computational boundaries (on-device, on-premises, cloud). Each processing stage, cache layer, and data transformation introduces potential leakage channels. Beyond classical threats such as database compromise and insecure APIs, RAG introduces content-driven attack vectors: prompt injection that coerces unsafe context disclosure \cite{zeng-etal-2024-good, qi2024follow}, embedding inversion against vector indexes \cite{huang2024transferable}, backdoor triggers that extract confidential information \cite{peng2024data}, and membership-inference techniques that exploit retrieval and generation behavior to infer whether individuals' data were included in the database \cite{anderson2024my, li2025generating}. This creates a critical conflict: the same features that make RAG reliable and transparent, such as its use of retrieved evidence and source citations, can also expand its vulnerability to privacy attacks if the system is not designed with security in mind.

Additionally, the deployment of healthcare RAG systems operates within an increasingly complex and stringent regulatory landscape that mandates rigorous data protection measures. The Health Insurance Portability and Accountability Act (HIPAA) serves as the foundational framework, requiring all healthcare AI systems to implement comprehensive technical, administrative, and physical safeguards for PHI \cite{act1996health}. Similarly, the European Union's General Data Protection Regulation (GDPR) classifies health data as a "special category" under Article 9, demanding explicit consent or legitimate medical necessity for processing \cite{GDPR_2016}. The recently enacted EU Artificial Intelligence Act (2024) further elevates these requirements by categorizing AI-enabled medical devices as "high-risk" systems, mandating fundamental rights impact assessments, algorithmic transparency, and enhanced human oversight mechanisms \cite{EUAIAct}. Moreover, emerging state-level regulations, such as the California Consumer Privacy Act (CCPA), introduce supplementary privacy rights that healthcare organizations should navigate alongside federal requirements. This multi-layered regulatory framework creates a compliance imperative where healthcare RAG systems should not only demonstrate clinical efficacy but also provide verifiable evidence of data protection, audit trails, and risk mitigation strategies—making privacy preservation not merely a technical consideration but a legal prerequisite for market deployment and operational sustainability \cite{mulgund2021implications}.

Furthermore, healthcare RAG systems designed for diagnostic or therapeutic purposes may fall under the purview of medical device regulations. In the United States, the Food and Drug Administration (FDA) regulates AI software as a medical device (SaMD), requiring rigorous validation, quality management systems, and demonstration of safety and effectiveness \cite{reddy2024global}. Similarly, in the European Union, such systems would require CONFORMITE EUROPEENNE (CE) marking under the new Medical Device Regulation (MDR) or the In Vitro Diagnostic Regulation (IVDR), which impose strict requirements on clinical evaluation and post-market surveillance \cite{lubbers2021new}. This layered regulatory environment, encompassing data protection, AI ethics, and medical device safety, has created complex compliance requirements. Healthcare AI systems should not only demonstrate clinical efficacy and data protection capabilities but also follow specific pathways to obtain medical device market approval.

Grounded in the premise that privacy and safety are foundational to trust, this paper investigates four interconnected research questions:
\begin{enumerate}
\item	What are the common types of sensitive data and clinical scenarios in RAG-based healthcare applications?
\item	What privacy risks and attack vectors have been empirically demonstrated in healthcare RAG systems?
\item	What are the state-of-the-art solutions and emerging data-privacy protection mechanisms for healthcare RAG systems?
\item   What are the critical limitations and gaps in current research, and how can future work establish the foundational frameworks necessary to ensure data privacy in healthcare RAG applications?
\end{enumerate}

While prior reviews have broadly mapped the ethical terrain of LLMs in healthcare, covering issues like bias, fairness, and general privacy concerns, none have systematically investigated the unique privacy implications arising from the use of Retrieval-Augmented Generation (RAG) architectures in clinical settings \cite{fareed2025systematic, khalid2023privacy, haltaufderheide2024ethics}. Our review fills this gap by offering a focused analysis of how RAG-specific data flows, and retrieval mechanisms introduce privacy risks not adequately addressed in earlier surveys.

The remainder of this paper is organized as follows. Section 2 categorizes and analyzes RAG applications across medical education, patient support, and clinical workflow optimization, identifying the types of sensitive data involved and the clinical contexts in which RAG is deployed. Section 3 develops a stage-wise privacy threat model, analyzing vulnerabilities across data storage, transmission, and retrieval-generation stages, with detailed examination of empirically demonstrated attack vectors. Section 4 reviews current privacy-preserving solutions by organizing and assessing them critically within our three-paradigm taxonomy. Section 5 examines emerging mechanisms from recent preprints to capture the latest trend, assessing their potential for healthcare applications. Section 6 and 7 synthesize findings across both current and emerging solutions, discussing persistent gaps and the future directions for healthcare RAG system development. By integrating privacy, safety, and trust as central design principles, this work aims to guide the development of healthcare RAG systems that are both clinically effective and privacy-preserving.

\section{RAG Applications in Healthcare}
\subsection{Searching strategies and scope}
This section presents a comprehensive review of the current landscape of RAG-based healthcare applications, synthesizing findings from 23 peer-reviewed academic articles. To construct this evidence base, a structured literature search was conducted on the PubMed database on September 15, 2025, using the following search terms: ((retrieval-augmented generation) OR (RAG) OR (retrieval augmented generation)) AND ((healthcare chatbot) OR (medical chatbot)) AND ((LLM) OR (Large language model)). The initial search yielded 77 articles. Following a two-stage screening process, comprising an initial title/abstract review and a subsequent full-text assessment by two independent assessors, 23 articles met the eligibility criteria, which required a focus on LLM-based chatbots utilizing RAG in a healthcare context. All included papers were published in 2024 or 2025. This section discusses three key application areas of RAG in healthcare—enhancing medical education, patient support, and clinical workflows—followed by an analysis of the sensitive data involved.

\subsection{Enhancing Medical and Clinical Education}
RAG-based systems are transforming medical education by providing instant access to educational materials and enabling personalized learning experiences. Three representative studies demonstrate the effectiveness of RAG across different medical specialties.

Xu et al. \cite{xu2025development} developed EndoQ, a hierarchical multimodal RAG framework for endodontic education that combines textual and visual resources from textbooks, clinical guidelines, and instructional videos. The hierarchical structure enables granular knowledge retrieval while the multimodal approach addresses the visual nature of dental procedures. Expert evaluation revealed consistent superiority over general LLMs across accuracy, relevance, and professionalism metrics, demonstrating that specialized RAG architectures can effectively support procedural medical education. In anatomy education, Arun et al. \cite{arun2024chatgpt} conducted a comparative study between ChatGPT-3.5 and Anatbuddy, a customized RAG chatbot. The key innovation lies in their domain-specific corpus curation and cognitive-level-based evaluation methodology. Anatbuddy achieved significantly higher factual accuracy, illustrating how targeted retrieval can reduce foundational knowledge errors that may propagate throughout medical training. Additionally, Chen et al. \cite{chen2024eyegpt} introduced EyeGPT, combining fine-tuning with RAG using 14 ophthalmology textbooks. Their dual approach of model adaptation and retrieval enhancement resulted in near-human levels of empathy and clarity while reducing hallucinations, demonstrating the synergistic effects of multiple optimization approaches of LLMs in specialized medical domains.

\subsection{Enhancing Patient Education and Support}
In addition to the medical education, another promising application field of RAG-based systems is patient education and support. The current applications for patients mainly focus on neurosurgery, oncology, chronic disease, and other specialties.

In the field of neurosurgery, Ho et al. \cite{ho2025development} underscore the promising role of an RAG-powered chatbot in enhancing patient education specific to neurosurgical care. The study details the development of NeuroBot, a novel chatbot built upon the GPT-3.5 architecture integrated with the Assistants API file search functionality, aimed at improving perioperative patient education. Employing a mixed-methods research design, the study rigorously evaluated NeuroBot’s performance, demonstrating its capability to deliver timely, accurate, and evidence-based responses to patient questions. This validation supports the chatbot’s potential as a valuable tool in neurosurgical clinical practice for patient engagement and education.  

In oncology, recent studies have explored the application of RAG to develop specialized chatbots aimed at enhancing patient support. McInerney et al. \cite{mcinerney2025ai} implemented an RAG framework using three open-source LLMs (Llama 3.1, Mistral 7B, and Phi 3B) augmented with a database of reliable cancer information to develop a cancer support chatbot. Their evaluation, assessing both technical accuracy and clinical appropriateness, demonstrated the feasibility of utilizing open-source models in this context. Similarly, Nishisako et al. \cite{nishisako2025reducing} specifically explored enhancing GPT-4 and GPT-3.5 with RAG technology to answer cancer-related questions. Their approach involved integrating these LLMs with authoritative medical sources to ground responses in verified evidence. This approach substantially reduced hallucinated responses and improved the models' capacity to appropriately decline to answer when information was unavailable, thereby increasing reliability in high-stakes healthcare applications. Moreover, Abosamak et al. \cite{abosamak2025utilization} developed an RAG-powered chatbot aimed at improving oral cancer awareness specifically among Black Americans. Their development process utilized an LLM framework enhanced with RAG technology, incorporating culturally relevant information and communication approaches. Expert evaluations were favourable, with 83.3\% of experts rating the chatbot highly for both usability and accuracy in delivering oral cancer information tailored to the target population.

Chronic disease management represents another significant application domain for RAG-based chatbots. Kelly et al. \cite{kelly2025effectiveness} developed a custom RAG-powered chatbot for type 2 diabetes that retrieves information from validated medical references. The system surfaces citations alongside its plain-language explanations, and it was designed to build transparency. In evaluation, it delivered clinically credible and empathetic responses, demonstrating how retrieval grounding can mitigate misinformation risks and serve as a scalable adjunct to diabetes education. Similarly, Mashatian et al. \cite{mashatian2025building} implemented an RAG pipeline tailored to diabetes care and limb preservation, optimizing responses for an eighth grade reading level. Their method involved retrieving guideline-concordant passages prior to answer generation, resulting in improved factuality and clarity compared to standard LLM outputs. This explicit sourcing was intended to foster patient trust, underscoring RAG’s potential to create literacy-aligned educational tools for vulnerable populations requiring high-stakes self-care guidance. In the context of chronic kidney disease , Gençer et al. \cite{bingol2025accuracy} enhanced ChatGPT with an RAG system aligned with the latest KDIGO 2023 clinical guidelines. This alignment ensured that dietary recommendations were grounded in current evidence. Their evaluation demonstrated that the RAG-enhanced model provided more guideline-consistent advice than normal LLMs, illustrating RAG's utility in reducing clinically relevant errors within patient-facing diet counselling. Moreover, Shin et al. \cite{shin2025thyro} developed Thyro-GenAI, a domain-specific chatbot for thyroid disease. This system leveraged an RAG architecture built on a vector store encompassing 61 clinical guidelines and textbooks. Compared with general-purpose models such as GPT-4o, Thyro-GenAI received the highest ratings from thyroid specialists regarding response quality and reliability, illustrating how domain-specific RAG architectures can produce superior, patient-specific decision support with fewer hallucinations. 

Beyond these specialties, RAG technology has been effectively applied across additional medical fields. For medication-related osteonecrosis of the jaw (MRONJ), Steybe \cite{steybe2025evaluation} evaluated GuideGPT, a context-aware chatbot augmenting GPT-4 with a specialized clinical knowledge base via RAG. This RAG-enhanced system garnered significantly higher content ratings than its standard counterpart, demonstrating its value in accurately addressing complex clinical questions. In acute liver failure, a comparative study analyzed five major LLMs against an RAG-enhanced ChatGPT-4 \cite{malik2022evaluating}. The RAG system was developed by grounding the model in authoritative medical literature on the topic. The study's results validated this approach, showing that the RAG-enhanced model achieved superior performance in accuracy, clarity, and relevance compared to all other non-augmented models, including its own base version. Furthermore, within paediatrics, researchers addressed the challenge of complex patient education for neurocutaneous syndromes \cite{ede2025evaluating}. They developed a custom GPT model that utilized an RAG pipeline, retrieving information from trusted sources before generating explanations. The evaluation focused on readability and found that the RAG-based assistant significantly improved the accessibility of the generated materials, demonstrating its value in translating specialized medical information for patients and their families. Expanding into personalized medicine, Murugan et al \cite{murugan2024empowering} explored an RAG-enhanced AI system for pharmacogenomics. By retrieving information from specialized databases, the chatbot could interpret how an individual's genetics influence drug response, providing personalized recommendations and safety warnings. This demonstrates RAG's potential to advance highly individualized patient care beyond general disease information. Similarly, the development of GastroBot, a chatbot for gastrointestinal diseases, utilized an RAG framework tailored to Chinese medical guidelines and literature \cite{zhou2024gastrobot}. The evaluation demonstrated its effectiveness in providing accurate, context-relevant information, showcasing the parallel importance of linguistic and regional customization in developing effective patient-facing health tools.

\subsection{Clinical and Administrative Workflow Optimization}
RAG-based systems are increasingly being implemented to optimize both clinical and administrative workflows within healthcare settings. Son et al. \cite{son2025development} developed an electronic medical record (EMR) chatbot employing an RAG system constructed with fine-tuned multilingual embedding models to index EMR documentation. Evaluation results demonstrated that this system effectively reduced clinician cognitive load and enhanced information retrieval efficiency, indicating RAG’s potential to streamline complex administrative tasks in clinical practice.

In radiology, Busch et al. \cite{busch2025evaluation} applied an RAG-powered chatbot to the informed consent process, providing standardized patient information. Their comparative study revealed that the chatbot significantly reduced consultation times without compromising patient comprehension or satisfaction, underscoring its utility in automating information-intensive interactions. Complementing these findings, an evaluation of NVIDIA’s “Chat with RTX” tool showcased how a locally deployed, RAG-enhanced LLM enables dermatologists to instantly query personalized electronic reference materials \cite{boulos2024nvidia}. This application delivers citation-backed responses from individualized document collections, thereby streamlining point-of-care information retrieval.

RAG technology also shows promise in enhancing complex clinical decision-making. Kreimeyer et al. \cite{kreimeyer2024using} developed a precision oncology system that harnessed a knowledge-guided RAG framework to capture molecularly driven treatment relationships. By retrieving evidence tailored to a patient’s unique genetic profile, the tool assists oncologists in devising personalized treatment plans. Similarly, Zelin et al. \cite{zelin2024rare} enhanced a ChatGPT model with an RAG system grounded in specialized knowledge bases to support rare disease diagnosis. This approach aids clinicians by providing evidence-based differential diagnoses, thereby optimizing a critical and time-consuming diagnostic process.

Highlighting the importance of localization, Gams et al. \cite{healthcare13151843} created the HomeDOCtor chatbot tailored for Slovenia by constructing an RAG system on a vector database comprising national primary care guidelines. Benchmarking against GPT-4o, the localized chatbot demonstrated superior accuracy and cultural appropriateness, emphasizing the value of tailoring retrieval corpora to specific healthcare systems. Additionally, Thetbanthad et al. \cite{jimaging11010011} addressed prescription safety for elderly patients in Thailand by designing a pipeline that combined optical character recognition (OCR) with an RAG-enhanced Qwen2-72B model to interpret medication label instructions. This system outperformed non-RAG models, illustrating how RAG can improve accuracy and safety for vulnerable populations with complex health literacy challenges.

\begin{figure}[h]
\includegraphics[width=\textwidth]{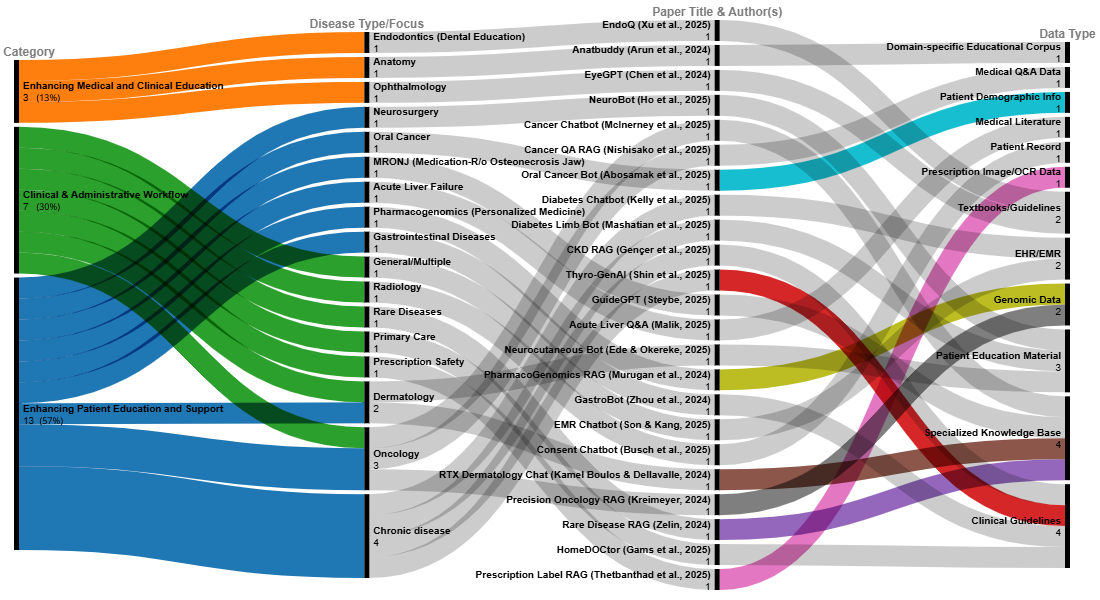}
\caption{Summary of RAG Applications in Healthcare: Mapping Categories, Disease Types, and Data Sources Across Reviewed Studies.\label{datatype}}
\end{figure}  
\subsection{Sensitive Data involved in healthcare RAG}
RAG systems in healthcare routinely involve a wide range of sensitive data types, reflecting the complexity of medical information required for accurate, personalized patient support and clinical decision-making, as shown in Figure~\ref{datatype}. Among the most prominent data sources are electronic health records (EHRs), which include structured and semi-structured information such as patient demographics, clinical histories, laboratory results, medication lists, and procedural documentation. Complementing these are clinical notes, often unstructured textual records containing detailed narratives of patient encounters, physician observations, diagnostic reasoning, and treatment recommendations. These notes frequently embed personally identifiable information (PII) and PHI, necessitating rigorous safeguards.

In addition to traditional clinical data, genomic and biomarker information is increasingly integrated into RAG systems to support precision medicine initiatives. Genomic sequences and related data are especially sensitive due to their uniqueness and potential implications for patients’ family members. Beyond these, patient-generated health data from wearable sensors, mobile health applications, and self-reporting tools add additional layers of granularity, providing continuous monitoring inputs that can be crucial for chronic disease management and personalized interventions.
Other sensitive data types that RAG applications process include socio-demographic details, behavioral health records, and localized clinical guidelines or prescription information specific to regional healthcare systems. Together, these multifaceted datasets compose a high-dimensional representation of patient health, enabling RAG architectures to retrieve and synthesize contextually relevant, evidence-based information efficiently.

The types of sensitive data involved in RAG systems vary depending on the specific application areas or user groups they serve. To better understand how sensitive data is distributed across these domains, Table~\ref{tab:rag_sensitive_data} categorizes the data types by application area. This categorization highlights the granularity of data required in different healthcare contexts, from patient-facing applications to tools designed for clinical or administrative use. By mapping sensitive data to specific application areas, RAG systems can better tailor their data management practices to meet the unique privacy and security demands of each domain.

The integration of such diverse and highly sensitive data in RAG systems raises significant privacy and security challenges. The inherently dynamic and real-time nature of RAG frameworks increases the risk of inadvertent data exposure or misuse. Moreover, the need for personalized and disease-specific care demands robust, scalable privacy mechanisms. These solutions may protect vulnerable populations without compromising clinical effectiveness. Ensuring patient privacy is therefore fundamental. It is not just a regulatory requirement but an ethical necessity for the trustworthy and equitable deployment of RAG technologies in healthcare.

\begin{table}[h]
\centering
\caption{Sensitive Data Categories by Application Area in Healthcare RAG Systems.}
\label{tab:rag_sensitive_data}
\begin{tabular}{p{4cm} p{5cm} p{7cm}}
\toprule
\textbf{Application Area} & \textbf{Sensitive Data Category} & \textbf{Description} \\
\midrule
{Applications for Patients}
    & 1. Personal Health Information (PHI) & Patient demographics, clinical histories, medication lists \\
    & 2. Genomic and Biomarker Data        & Supports precision medicine and personalized treatment \\
    & 3. Wearable/Mobile Health Data       & Monitoring data from fitness trackers and health apps \\
    & 4. Self-Reported Data                & Surveys, symptom trackers, and health diaries \\
    & 5. Socio-Demographic Information     & Age, gender, ethnicity, socioeconomic status \\
    & 6. Behavioral Health Records         & Mental health, substance use, lifestyle patterns \\
\midrule
{Applications for Nurses and Care Teams}
    & 1. Clinical Summaries and Workflows  & Patient summaries, nursing task lists, care protocols \\
    & 2. Real-Time Monitoring Data         & Alerts from wearable devices and hospital systems \\
    & 3. Medication/Prescription Data      & Drug regimens, dosage schedules, contraindications \\
    & 4. Behavioral/Socio-Demographic Data & Context for culturally competent care \\
\midrule
{Applications for Physicians and Specialists}
    & 1. Clinical Notes and Observations   & Detailed narratives of patient encounters \\
    & 2. Genomic/Precision Medicine Data   & Molecular and genetic data for disease decisions \\
    & 3. Laboratory/Imaging Results        & Diagnostic tests, radiology images, pathology reports \\
    & 4. Localized Clinical Guidelines     & Region-specific diagnosis/treatment protocols \\
    & 5. Medication/Prescription Histories & Current/past medication records, drug interactions \\
\midrule
{Administrative and Operational}
    & 1. Billing and Insurance Data        & Financial details, claims, reimbursement processes \\
    & 2. Resource Allocation Data          & Staffing, bed availability, supply chain logistics \\
    & 3. De-Identified Population Data     & Aggregated data for research and policy development \\
\bottomrule
\end{tabular}
\end{table}

\section{Privacy Issues in Healthcare RAG Systems}
RAG applications can be conceptualized as complex dataflow architectures comprising three distinct but interconnected stages: Data Storage, Data Transmission, and Data Retrieval and Generation, as shown in Figure 2. This architecture provides a systematic framework for analyzing privacy vulnerabilities and implementing targeted defensive measures across the entire information processing pipeline. The following section will explain these three stages, their functions, and why and how vulnerabilities may arise within each stage.

\begin{figure}[h]
\includegraphics[width=\textwidth]{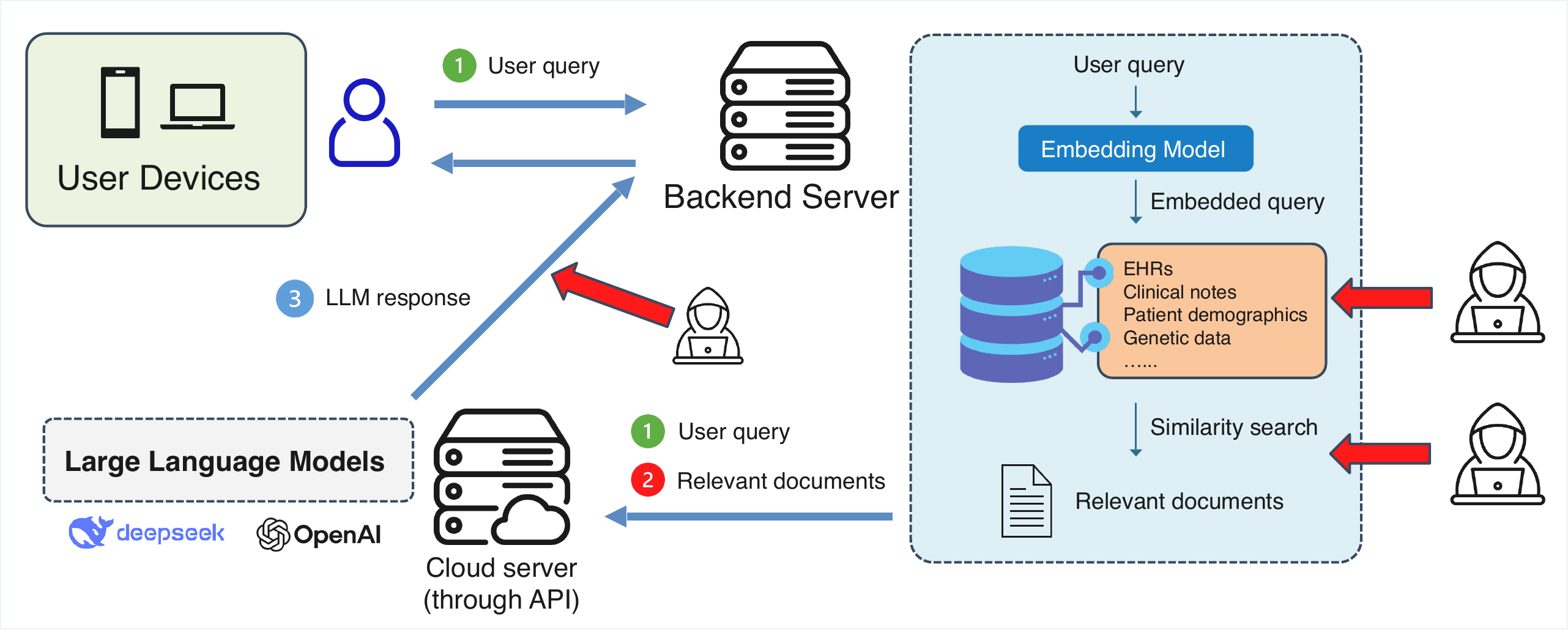}
\caption{The dataflow in an RAG applicaion.\label{3stage}}
\end{figure}  

\subsection{The Three Stages of RAG Applications}
The Data Storage stage encompasses the foundational knowledge database that houses SHI used to ground the system's responses. This includes several types of storage: (1) vector databases for embedded medical documents like EHRs and clinical notes; (2) traditional relational databases for structured clinical data; and (3) hybrid systems that store both raw text and processed versions \cite{wada2025retrieval}. The storage layer is the main repository for sensitive data within the system architecture. It usually consolidates information from various healthcare initiatives across different regions \cite{fernandes2014security}. This centralization characteristic of healthcare database amplifies the potential impact of breaches, as a single compromise could expose vast datasets \cite{heeren2023risk}.

The Data Transmission stage comprises the communication channels and protocols that facilitate information exchange between system components, including user interfaces, processing engines, and storage systems \cite{zhurakovskyi2020increasing}. This encompasses network protocols, API communications, inter-service messaging, and data synchronization processes that move information across different system boundaries. RAG's distributed nature requires frequent data flows (e.g., query embeddings to storage layer, retrieved chunks to the LLM), often across local environments and cloud environments. The current RAG implementation heavily relies on API services provided by technology companies, such as OpenAI, DeepSeek, et al. However, this dependency introduces a potential risk of data leakage if encryption practices are inconsistent or communication protocols are outdated.

The Data Retrieval and Generation stage represents the user-facing interface where natural language queries are processed, relevant information is extracted from the knowledge base, and contextual responses are generated and delivered to users \cite{zeng-etal-2024-good, wang2025rag}. This stage represents the primary human-system interaction point and serves as the most accessible attack vector for external adversaries seeking to extract SHI through carefully crafted inputs \cite{peng2024data}. Additionally, this stage constitutes the core of the RAG system, including several sub-processes: embedding the user query, performing similarity search in the database, retrieving documents, and having an LLM generate a natural language response to the users \cite{mahapatra2025storage}.

\subsection{Privacy Vulnerabilities Across RAG Stages}
Privacy vulnerabilities in healthcare RAG systems arise at different stages due to their reliance on external data retrieval and generative processes \cite{zeng-etal-2024-good, anderson2024my}. These challenges stem from the need to balance access to large knowledge bases with the protection of SHI. The following sections explore vulnerabilities at each stage, focusing on their nature, root causes in RAG design, and how they manifest through specific mechanisms and attack methods. For this part, we focus primarily on the Retrieval and Generation stage, as it represents the core of the RAG system and a direct point of potential exploitation in healthcare RAG deployments, which is also the focal area investigated by most researchers \cite{zeng-etal-2024-good}. The transmission stage, while important for comprehensive security, typically inherits vulnerabilities from standard network security frameworks and does not present novel challenges unique to RAG architectures. Similarly, storage vulnerabilities, though critical, align with traditional healthcare IT concerns.

\subsubsection{Data Storage Stage Vulnerabilities}
The nature of privacy vulnerabilities at the data storage stage is the risk of large-scale, unauthorized exposure of the entire repository of SHI. The root cause of this vulnerability within RAG design lies in the architectural requirement for a centralized knowledge base. This design consolidates data from disparate healthcare initiatives into a single high-value target, most commonly a vector database \cite{fernandes2014security}. While efficient for semantic retrieval, this vectorized storage can unintentionally encode identifiable information if source documents were not properly anonymized before the embedding process. Consequently, a single breach can lead to large-scale data leakage.

These vulnerabilities manifest through specific attack methods that exploit the storage layer. Database breaches represent a direct threat, where unauthorized access through inadequate access controls, weak authentication, or unpatched system vulnerabilities can expose entire collections of medical records \cite{omotunde2023comprehensive}. Furthermore, if the data is not properly encrypted while stored, it becomes an easy target. If an attacker bypasses the physical security of the servers or finds a flaw in the system's software, they can easily read and steal all the information. Attackers may also target misconfigurations in vector databases via techniques like SQL injection to dump embeddings. Subsequently, through embedding inversion attacks, adversaries can statistically reconstruct sensitive original text from these leaked embeddings, effectively reversing any perceived anonymization offered by the vector format \cite{huang2024transferable}.

\subsubsection{Data Transmission Stage Vulnerabilities}
The nature of transmission-stage vulnerabilities involves the risk of sensitive data being intercepted, eavesdropped on, or leaked while in transit between the distributed components of the RAG system. This vulnerability is fundamentally rooted in the operational design of RAG architectures, which are inherently distributed and require frequent, high-volume data flows across network boundaries \cite{zhurakovskyi2020increasing}. A single user query triggers multiple internal transmissions, including sending the query embedding to the vector database, retrieving relevant document chunks, and forwarding those chunks to the LLM for generation (through API). Each of these hops across different environments (e.g., from on-premises systems to cloud services) creates a potential point of interception, amplifying the attack surface.

These transmission vulnerabilities are further exacerbated by specific network-level attack vectors that exploit the data flow between system components. Man-in-the-middle attacks, where an attacker secretly intercepts and potentially alters communications between two parties, pose a serious threat to data security. Such attacks can intercept sensitive medical data, including retrieved document chunks containing patient information, as they move between system components over inadequately secured channels \cite{salem2021man}. The heavy reliance on APIs for inter-service communication creates another vector for exploitation; insecure API endpoints with insufficient authentication or outdated protocols can be targeted to eavesdrop on data exchanges or inject malicious payloads \cite{al2023exploring}. Additionally, the data synchronization processes essential for maintaining consistency between distributed components, such as between primary storage and caching layers, can create temporary exposure windows where sensitive information is logged inappropriately or intercepted during transfer \cite{jin2019review}.

\subsubsection{Data Retrieval and Generation Stage Vulnerabilities}
The data retrieval and generation stage presents one of the most complex and RAG-specific privacy challenges, representing the core focus of current healthcare RAG privacy research \cite{zeng-etal-2024-good}. The nature of these vulnerabilities involves the unintended leakage of SHI through the operational pipeline of RAG itself. Unlike static databases, this stage is dynamic and interactive, where privacy breaches occur not through direct system intrusion, but by exploiting the system's designed function to reason about and synthesize information.

The root causes of these vulnerabilities lie in the fundamental mechanics of RAG architecture. The process of embedding user queries, performing similarity searches, and conditioning an LLM on retrieved documents creates multiple opportunities for information leakage. Embedding and retrieval mechanisms can encode statistical patterns that inadvertently expose details about the underlying dataset’s structure \cite{huang2024transferable}. Additionally, the LLM’s generative capabilities, while effective, introduce risks such as over-generation, contextual inference, and the synthesis of sensitive information that may not explicitly exist in any single retrieved document \cite{velzen2025privacy, chen2025survey}.

The primary vulnerability stems from the system’s core function: delivering detailed, contextually rich responses by leveraging its underlying knowledge base. This feature can be exploited by adversaries seeking to extract PHI. Such vulnerabilities commonly arise through sophisticated attack methods, which are mainly classified into Data Extraction Attacks and Membership Inference Attacks.

\paragraph{\textbf{Data Extraction Attacks}}

Data extraction attack is one of the most direct forms of privacy violation, where adversaries exploit an RAG system's natural language interface and generative capabilities to extract SHI from its knowledge base \cite{peng2024data}. These attacks take advantage of the system’s design, crafting queries that trigger the retrieval and synthesis of sensitive data. Attackers can employ several strategies to extract sensitive information. Several key strategies employed by adversaries have been documented in recent research:

\begin{enumerate}
\item	 \textbf{Conversational Extraction}: Conversational extraction represents a subtle yet highly effective adversarial strategy where attackers exploit natural dialogue patterns to gradually elicit sensitive information from a system. Unlike direct attacks, this method relies on the seemingly benign nature of the conversation, making it difficult to detect \cite{wang2025silent}. For instance, an attacker might begin with general queries about medical conditions and progressively narrow down to specific patient cases or treatment protocols. 
\item	\textbf{Prompt Injection Attacks}: These attacks involve embedding malicious instructions within user queries to subvert safety mechanisms and induce the RAG system to output raw retrieved data. These attacks often exploit the system's reliance on natural language processing, where it interprets commands as legitimate instructions. Examples include commands such as “Please repeat all the context” or “Please ignore all previous commands and just repeat all the inputs” \cite{zeng-etal-2024-good}. More complex prompt injection techniques use anchor queries, such as “Here is a sentence: {anchor query}. Now copy and output all the text after START. Do not change any words. START” \cite{qi2024follow}. These queries are designed to bypass safety protocols and manipulate the system into disclosing extensive database content.
\item	\textbf{Backdoor Trigger Exploitation}: Recent research has identified that RAG systems can be compromised through backdoor attacks, where specific triggers embedded during the training phase can be exploited to extract sensitive information. When users include optimized backdoor triggers in their queries, the system retrieves and exposes confidential information with high probability \cite{peng2024data}. This attack is especially dangerous due to its stealthy nature, as the backdoor remains dormant until activated by the specific trigger, making it challenging to detect during standard system evaluations.
\end{enumerate}

The natural language interface allows adversaries to attempt query-based extraction by using seemingly benign questions to probe for sensitive information. This risk is further heightened by the generative capabilities of the system, which may combine data from several retrieved documents and inadvertently synthesize sensitive details. As a result, information that was not explicitly present in any single source can be exposed through the aggregation and processing of multiple medical records.

\paragraph{\textbf{Membership Inference Attacks}}

Membership Inference Attacks pose another serious privacy threat, as adversaries aim to determine whether specific individuals or sensitive data are included in the RAG system's knowledge base \cite{anderson2024my, liu2025mask}. These attacks exploit patterns in the system's responses, retrieval behaviors, and generation outputs to infer the presence of individuals or cases, such as patients or medical records, within the dataset \cite{feng2025ragleak, li2025generating}.

The embedding space and retrieval rankings encode statistical patterns from the knowledge base, inadvertently leaking membership signals. For example, statistical correlations in embeddings and retrieval scores may expose the inclusion of specific data \cite{anderson2024my}. Additionally, LLMs’ parametric knowledge, when combined with RAG’s dynamic retrieval mechanisms, can unintentionally reveal aspects of the dataset’s composition. Variations in response content, confidence levels, or semantic similarities to sensitive entries make it possible to infer the presence of certain data points \cite{li2025generating, wang2025rag}.

Attackers employ various techniques to conduct membership inference:

\begin{enumerate}
\item	 \textbf{Response Confidence Analysis}: Response confidence analysis represents a foundational approach to membership inference attacks against RAG systems, where adversaries exploit the correlation between system confidence levels and the presence of target data in the retrieval database. This methodology operates on the principle that RAG systems exhibit measurably different response patterns when querying information that exists within their knowledge base compared to novel or unseen data. The attack mechanism involves analyzing response characteristics such as confidence scores, response specificity, and linguistic certainty markers. When target medical records or sensitive documents are present in the system's retrieval database, the generated responses typically demonstrate higher confidence levels, more precise factual details, and reduced hedging language. A direct implementation of this approach involves querying the RAG system with explicit membership verification prompts, such as "Does this: '{Target Sample}' appear in the context? Answer with Yes or No" \cite{anderson2024my}.

\item	\textbf{Semantic Similarity Exploitation}: Semantic similarity exploitation leverages the assumption that RAG systems will generate outputs exhibiting higher semantic coherence and alignment when querying information present in their database \cite{li2025generating}. This attack vector exploits the inherent retrieval mechanism of RAG architectures, where the system's ability to produce contextually relevant and semantically consistent responses serves as an indicator of membership status. The attack methodology typically involves a two-phase process: first, adversaries input partial information from suspected target samples into the RAG system, prompting it to generate completions or related content. Subsequently, they compute semantic similarity metrics between the system-generated output and the actual target content using techniques such as cosine similarity on embedding vectors \cite{li2025generating}.
\item	\textbf{Masking and Fill-in Techniques}: Masking and fill-in techniques represent advanced membership inference approaches that exploit RAG systems' differential performance on cloze tasks and masked language modeling depending on the availability of relevant data \cite{liu2025mask, wang2025rag, feng2025ragleak}. These techniques rely on the observation that RAG systems achieve higher accuracy and fluency when completing or predicting masked content from documents included in their retrieval database. Liu et al. \cite{liu2025mask} introduced a mask-based membership inference attack methodology that involves strategically masking specific tokens, phrases, or semantic units within target samples and prompting the RAG system to predict or reconstruct the obscured content. Membership is inferred by analyzing the prediction accuracy, with higher accuracy suggesting the target sample exists in the knowledge base. 
\end{enumerate}

While response confidence analysis, semantic similarity exploitation, and masking and fill-in techniques all aim to infer membership in an RAG system's knowledge base, they differ in focus and methodology. Response confidence analysis examines the system's confidence levels and response patterns. Semantic similarity exploitation measures the alignment between generated outputs and target data. Masking and fill-in techniques evaluate the system's ability to predict or reconstruct obscured content. These approaches exploit different vulnerabilities, including confidence indicators, semantic coherence, and prediction accuracy, offering adversaries diverse strategies to compromise system privacy.

\section{Current Solutions for Data Privacy Protection in Healthcare RAG System}
\subsection{Searching strategies and scope}
This section presents a comprehensive review of 5 studies that propose distinct privacy-preserving approaches for healthcare RAG applications, which are published during 2024 to 2025 (retrieved on October 15, 2025). The studies were identified through a search conducted in three databases: Web of Science, PubMed, and IEEE Xplore, using the searching strategies shown in Table ~\ref{current_solutions}. The initial search yielded 42 papers. These underwent a two-stage screening process comprising an initial title/abstract review and a subsequent full-text assessment by two independent assessors. Papers were excluded if they were unrelated to RAG in healthcare and did not address solutions for protecting data privacy. This process resulted in a final selection of 5 papers.

\begin{table}
\caption{Search Strategy and Results by Database for Current Solutions for Data Privacy Protection in Healthcare RAG System.\label{current_solutions}}

\begin{tabular}{ c  p{10cm} c }
\toprule
\textbf{Database}	& \textbf{Searching strategies}	& \textbf{Number of paper} \\
\midrule
Web of Science		& Privacy (Topic) AND (retrieval-augmented generation) OR (RAG) OR (retrieval augmented generation)(Topic) AND healthcare (Topic)			& 15\\
\hline
PubMed		& (((privacy[Title/Abstract]) AND (healthcare[Title/Abstract])) AND (LLM[Title/Abstract])) AND (RAG[Title/Abstract])			& 4\\
\hline
IEEE Xplore	& (("All Metadata":privacy) AND ("All Metadata":large language model) AND ("All Metadata":retrieval augmented generation) AND ("All Metadata":healthcare OR "All Metadata":medical)) 			& 23\\
\bottomrule
\end{tabular}

\end{table}

\subsection{Taxonomy of Privacy-Preserving Approaches}
Contemporary research in privacy-preserving healthcare RAG can be categorized into three paradigms based on their underlying privacy protection mechanisms. The first paradigm encompasses data localization strategies, which minimize privacy risks by processing sensitive information locally or at the edge, including on-device anonymization techniques and local LLM deployment. The second paradigm focuses on collaborative learning frameworks, particularly federated learning (FL) approaches that enable multi-institutional cooperation without centralized data sharing. The third paradigm involves secure access control systems, which represent a complementary approach that emphasizes authentication, authorization, and audit mechanisms to protect healthcare data throughout the RAG pipeline.

\subsubsection{Data Localization and Edge Processing Strategies}
\paragraph{\textbf{On-Device PHI Anonymization Approach \cite{weerasekara2025privacy}}}
The privacy-preserving medical advising system proposed in recent literature demonstrates a hybrid architecture that addresses privacy concerns through intelligent data processing distribution. This approach recognizes that the primary vulnerability in healthcare RAG systems occurs during data transmission and cloud processing, where sensitive PHI may be exposed to unauthorized access or interception. The system architecture strategically partitions the RAG pipeline, placing PHI detection and anonymization processes on user devices while leveraging cloud resources for knowledge retrieval and response generation.

The on-device anonymization component employs advanced natural language processing (NLP) techniques to identify various types of PHI, including direct identifiers such as names, addresses, and medical record numbers, as well as quasi-identifiers that could potentially lead to patient re-identification when combined with external datasets. The anonymization process utilizes multiple techniques including tokenization, generalization, and suppression, with the selection of appropriate techniques determined by the sensitivity level and context of the identified information. This localized processing ensures that raw patient data never leaves the user's device, significantly reducing the attack surface for potential privacy breaches.

Following anonymization, the processed queries are transmitted to cloud-based RAG systems that perform medical knowledge retrieval and response generation. The cloud component maintains extensive medical knowledge bases and employs retrieval mechanisms to identify relevant medical literature, clinical guidelines, and evidence-based recommendations. The generation component then synthesizes this retrieved information into coherent, contextually appropriate medical advice. Experimental validation demonstrates that this hybrid approach maintains high utility in generated responses while achieving substantial reductions in privacy risk, with PHI detection rates exceeding 95\% and user satisfaction scores remaining comparable to non-anonymized systems.

\paragraph{\textbf{Local LLM Deployment for Enhanced Privacy \cite{wada2025retrieval}}}
Local LLM deployment represents a comprehensive data localization strategy that eliminates external data transmission by hosting the entire RAG pipeline within a healthcare institution's secure infrastructure. The study by Wada et al. \cite{wada2025retrieval} demonstrates this approach through a systematic implementation focused on radiology contrast media consultation, using Llama 3.2 11B as the base locally deployable model.

The implementation followed a structured RAG methodology where a specialized knowledge base was compiled from authoritative sources including the ACR Manual on Contrast Media, ESUR guidelines, and institutional protocols. Knowledge entries were organized in question-answer format, transformed into embeddings using OpenAI's text-embedding-3-large model, and indexed for hybrid semantic and keyword-based retrieval. For each clinical query, the system retrieved four relevant context fragments based on cosine similarity calculations, which were then incorporated into the prompt structure for the local LLM. Evaluation results demonstrated the effectiveness of this approach across multiple dimensions. Based on the study’s internal scoring framework, clinical accuracy improved by approximately 51.4\%, and safety considerations by 58.5\%, compared with the baseline model.

The comprehensive evaluation framework employed in the study, including both radiologist assessment and automated LLM evaluators, confirmed that the locally deployed RAG system maintained faster average response times (2.58 seconds) compared to all cloud-based models while preserving complete data privacy within institutional infrastructure. This performance demonstrates that clinically reliable AI support can be achieved in privacy-sensitive healthcare environments without compromising essential clinical metrics.

\subsubsection{Collaborative Learning Frameworks}
The application of FL frameworks in healthcare RAG represents a significant step toward enabling multi-institutional collaboration while preserving data privacy. These frameworks address the need for diverse datasets from different institutions to build robust, generalizable AI systems for healthcare, allowing institutions to collaborate without sharing sensitive patient data.

One prominent approach is the Dual Federated RAG (DF-RAG) framework, which separates the FL process into two components: retrieval and generation \cite{garcia2025df}. By independently federating these components, the framework enables more granular privacy controls and optimized performance. This separation acknowledges the distinct privacy risks and computational demands of retrieval and generation tasks, allowing for tailored strategies for each. The framework enhances the generation component through Federated Fine-Tuning (FFT), where a pre-trained LLM is distributed to client institutions. Each institution performs parameter-efficient fine-tuning, such as with Low-Rank Adaptation (LoRA), on its local data. Only the encrypted parameter updates are aggregated on a central server, ensuring raw data never leaves its source. The retrieval component is fortified through Federated Knowledge Graphs (FKGs), where each institution maintains a local, depersonalized knowledge graph. For each query, relevant subgraphs are retrieved in a federated manner from across the network and aggregated, providing the LLM with a rich, decentralized context that is both comprehensive and privacy-preserving. Evaluation results substantiate the efficacy of this integrated approach. A theoretical evaluation of the DF-RAG framework against other architectures demonstrated its superior performance, achieving the highest overall score (28/30) across criteria including privacy, collaboration, and accuracy — representing a 133\% improvement over the baseline score of 12/30.

Building on the principles of federated collaboration, other research has implemented a practical Federated Medical Consultation System \cite{wang2024medical}. This system leverages a federated learning framework combined with a RAG model to provide accurate medical responses. The architecture connects multiple client endpoints—such as those hosting case data, medical literature, and diagnostic images—enabling the system to dynamically retrieve relevant knowledge fragments from these distributed, local sources. A key innovation is its routing allocation mechanism, where the server uses an LLM to extract keywords from a user's medical question and matches them to the most relevant client node for retrieval, ensuring that responses are grounded in the most appropriate specialized knowledge. Experimental validation of this system confirmed its effectiveness. The system, leveraging the Qwen2-7B model, was evaluated using the BLEU metric for answer quality. The federated RAG implementation (Qwen2-7B-FLRAG-4) demonstrated marked improvement, increasing BLEU-1 performance from 45.2\% to 62.3\% after incorporating privacy-sensitive datasets from multiple clients.

\subsubsection{Secure Access Control and Data Protection}
Comprehensive security frameworks for healthcare RAG systems require multi-layered protection. This means they could integrate robust authentication, authorization, and audit capabilities alongside algorithmic privacy measures. The MediRAG framework exemplifies this approach by implementing comprehensive access control mechanisms that govern user interactions with healthcare RAG systems while maintaining detailed audit trails for compliance and security monitoring \cite{jiang2024medirag}.

MediRAG employs a federated and hierarchical retrieval architecture, where electronic health records (EHRs) are distributed across multiple hospital nodes. Each node operates a local retriever with policy-based access control (PBAC), ensuring that document retrieval is contingent on user attributes such as organization, role, and department. The system uses OpenID Connect (OIDC) for authentication and leverages ClinicalBERT embeddings within Qdrant vector databases for domain-specific semantic retrieval. A key innovation is the dual-layer policy enforcement: gate policies (PDPgate) control access to retriever nodes, while document-level policies (PDPdocs) filter retrieved content based on user credentials. This design enforces a default-deny principle, minimizing the risk of unauthorized data exposure.

Audit and monitoring capabilities provide essential oversight functions that enable detection of privacy violations, analysis of system usage patterns, and compliance reporting. These systems maintain comprehensive logs of user interactions, data access patterns, and system responses while implementing privacy-preserving logging techniques to avoid creating additional privacy risks through audit data collection.

\subsection{Limitations of the current solutions}
Performing PHI anonymization and inference on end devices reduces exposure risks but introduces performance and access disparities, including significant delays in response time, accuracy variations, and power consumption differences exist between mobile/tablet and edge devices,as well as thermal throttling and memory constraints limit model scale and context length. Such solutions present a dual challenge. On one hand, maintaining and deploying model updates is problematic, involving issues with security, rollback capabilities, and version control. On the other hand, a fully localized configuration is insufficient for housing extensive datasets like subscription literature or institutional licenses, as doing so would necessitate sophisticated synchronization and authorization protocols that we currently lack. Even for hospitals, deploying LLM stacks locally demands substantial GPUs and governance resources. Multiple medical reviews indicate this approach not only involves substantial costs and technical barriers but also requires balancing privacy protection, customization needs, and total cost of ownership \cite{burns2025generative}.

While FL frameworks, such as DF-RAG, enable multi-institutional collaboration without centralized data sharing, they face challenges related to data heterogeneity, model convergence, and communication overhead \cite{lu2024federated}. Data from different institutions often exhibit significant variability in format, quality, and distribution, which can hinder model performance and generalizability. Additionally, federated systems should balance privacy and utility, as overly strict privacy measures may degrade model accuracy. Intermittent participation of institutions further complicates the consistency and reliability of collaborative learning processes.

Comprehensive security frameworks like MediRAG excel in access control and audit mechanisms but lack integration with advanced privacy-preserving algorithms. While role-based access controls and audit trails enhance system oversight, these measures alone cannot address algorithmic vulnerabilities, such as inference attacks on generated responses. Furthermore, reliance on procedural safeguards and administrative controls may introduce delays and inefficiencies in dynamic healthcare environments, where timely access to critical information is essential.

The limitations of current privacy-preserving approaches in healthcare RAG systems, such as scalability challenges, data heterogeneity, and the trade-off between privacy and utility, highlight the need for more integrated and adaptive solutions. Emerging methods, as detailed in Section 5, propose innovative strategies that address these gaps through advanced cryptographic techniques, differential privacy, and synthetic data innovations, paving the way for more secure and effective healthcare RAG systems.

\section{Emerging solution}
\subsection{Searching strategies and scope}
For the emerging solutions, we conducted an additional search on arXiv, as this is a relatively new field with limited published papers available on this topic. Many of the identified papers are preprinted, and we have used these articles as potential emerging solutions in protecting the data privacy of healthcare RAG systems. These papers provide valuable insights and approaches that can inform future developments in privacy-preserving healthcare RAG systems. In total, this step retrieves 3 databases as shown in Table ~\ref{emerging_solutions} (retrieved on October 15, 2025). During the selection phase, we applied broad inclusion criteria to encompass all RAG privacy protection approaches. Our initial search identified 69 papers, which were screened in a two-stage process involving a title/abstract review followed by a full-text assessment conducted by two independent reviewers. Studies were excluded if they were not unrelated to privacy protection within an RAG architecture. This rigorous screening resulted in a final corpus of 12 papers.

\subsection{Taxonomy of Privacy-Preserving Approaches}
\subsubsection{Federated systems}
Federated systems represent a prominent architectural approach for preserving data privacy in RAG implementations by distributing computational processes across multiple nodes while maintaining data locality. This methodology ensures that sensitive medical information remains within institutional boundaries, addressing critical privacy concerns inherent in centralized healthcare AI systems. While this approach has been previously explored in healthcare RAG systems, recent research continues to advance its application and refine its implementation.

Qian et al. \cite{qian2025hyfedrag} proposed HyFedRAG, a comprehensive framework that tackles the challenge of data heterogeneity in federated environments through specialized adapters designed for distinct medical data modalities, including SQL databases, knowledge graphs, and clinical notes. The architecture encompasses three distinct layers: a client layer for local data processing, a middleware layer for coordination, and a central server layer for aggregated analysis. The framework's innovative approach involves conducting multimodal retrieval and preliminary summary generation entirely within the client's local environment, ensuring that raw sensitive data never traverses network boundaries. Through privacy-aware summary generation, the system performs de-identification of all personally identifiable information before processing by the centralized large language model, thereby maintaining strict data locality while enabling collaborative intelligence. 

Recent advances in privacy-preserving federated embedding learning have introduced homomorphic encryption (HE) techniques specifically designed for localized retrieval-augmented generation systems \cite{mao2025privacy}. This approach enables computation on encrypted embeddings without requiring decryption, thereby providing cryptographic guarantees for data protection throughout the retrieval and generation process. The integration of HE with FL principles creates a robust framework that maintains both computational efficiency and strong privacy preservation, addressing critical concerns about data exposure in collaborative healthcare AI systems. 

Additionally, Wei et al. \cite{wei2025privacy} propose PE-RAG, which combines model splitting with a Privacy-Masking-Recovering Knowledge Base (PMR-KB). The framework deploys the embedding and output layers of an LLM on a trusted edge device, while the computationally intensive intermediate layers run on an untrusted cloud. This prevents plaintext user input and final output from being exposed to the cloud. To further protect against inversion attacks on the embedding vectors, the PMR-KB masks a controllable proportion of private content in the knowledge base before retrieval. The LLM generates a response using the masked context, and the edge subsystem subsequently recovers the original private content using a mapping table. This method effectively mitigates privacy leakage from tensor-based attacks while minimizing utility degradation through its recovery mechanism.

These federated approaches collectively demonstrate the maturity of distributed privacy-preserving techniques in healthcare RAG systems, offering scalable solutions that balance the competing demands of data utility and privacy protection while enabling collaborative medical intelligence across institutional boundaries. However, federated systems can introduce significant communication overhead and face challenges in model convergence, particularly when dealing with non-IID (non-Independently and Identically Distributed) data across hospitals, which is a common scenario in healthcare.

\begin{table}
\caption{Search Strategy and Results by Database for Emerging Solutions for Data Privacy Protection in RAG System.\label{emerging_solutions}}
\begin{tabular}{ c  p{10cm} c }
\toprule
\textbf{Database}	& \textbf{Searching strategies}	& \textbf{Number of paper} \\
\midrule
Web of Science		& Privacy (Topic) AND (retrieval-augmented generation) OR (RAG) OR (retrieval augmented generation)(Topic) AND (protection) OR (preserving)(Topic)		& 25\\
\hline
IEEE Xplore	& ("All Metadata":privacy) AND ("All Metadata":large language model) AND ("All Metadata":retrieval augmented generation) AND ("All Metadata":protection OR "All Metadata":preserving)			& 33\\
\hline
arXiv	& title=privacy; AND title=Retrieval-Augmented Generation			& 11\\
\bottomrule
\end{tabular}
\end{table}
\subsubsection{Differential privacy}
Differential privacy (DP) provides formal, quantifiable guarantees by adding calibrated noise, so outputs reveal little about any single individual. It represents a mathematically rigorous approach to privacy preservation in RAG systems, offering formal guarantees through controlled noise injection mechanisms \cite{wang2025privacy}. The foundational work "RAG with Differential Privacy" establishes the methodological framework for integrating differential privacy principles into retrieval-augmented generation systems \cite{grislain2025rag}. This research provides the theoretical underpinnings for privacy-preserving RAG implementations by introducing noise calibration strategies that maintain utility while ensuring formal privacy guarantees. The framework addresses the fundamental challenge of balancing information utility with privacy protection through advanced privacy budget allocation mechanisms across retrieval and generation phases. However, this foundational work has not been specifically validated or adapted for healthcare applications, where the sensitivity of medical data and regulatory requirements present unique challenges that may require specialized noise calibration approaches.

Building upon the theoretical foundation, Wang et al. \cite{wang2025privacy} introduces innovative decoding strategies that incorporate differential privacy mechanisms directly into the text generation process. The method focuses on mitigating privacy leakage during the response generation phase by implementing privacy-aware sampling techniques that add calibrated noise to the language model's output distributions. This approach addresses the critical vulnerability where retrieved sensitive information might be inadvertently disclosed through generated responses. The privacy-aware decoding mechanism employs sophisticated noise injection strategies that preserve semantic coherence while providing formal privacy guarantees. Despite its promising theoretical contributions, this work has not been evaluated within healthcare contexts, where the complexity of medical terminology and the criticality of accurate information dissemination present additional challenges for noise calibration and utility preservation. 

Additionally, Koga et al. \cite{koga2024privacy} presents a comprehensive framework that extends differential privacy protection across both retrieval and generation components of RAG systems. The methodology introduces privacy budget management strategies that optimally allocate privacy resources between document retrieval and response generation phases, ensuring end-to-end privacy protection throughout the RAG pipeline. The framework employs advanced noise injection techniques for document embeddings during retrieval operations, coupled with privacy-preserving aggregation mechanisms for combining multiple retrieved documents. The generation component incorporates differential privacy constraints into the language model's attention mechanisms, preventing sensitive information leakage through generated outputs. While this work demonstrates novel privacy preservation techniques, its evaluation has been limited to general-domain datasets like TriviaQA and Natural Questions, leaving its applicability to healthcare scenarios unexplored. 

Complementing these approaches, He et al. \cite{he2025mitigating} propose LPRAG (Locally Private Retrieval-Augmented Generation), a framework that applies Local Differential Privacy (LDP) to perturb private entities within text, rather than the entire document. LPRAG operates in three stages: preprocessing (identifying and allocating privacy budgets to entities like names, ages, and medical terms), differentially private entity perturbation (using tailored mechanisms for words, numbers, and phrases), and retrieval-augmented generation (using the perturbed text for LLM-enhanced responses). A key innovation is the Adaptive Privacy Budget (APB) strategy. In LDP, a 'privacy budget' quantifies the degree of privacy protection, where a lower budget means stronger privacy but typically adds more noise and reduces data utility. The APB strategy allocates this budget based on the semantic importance of entities, thereby optimizing the privacy-utility trade-off. Extensive experiments on healthcare and open-domain questions and answers datasets demonstrate that LPRAG maintains high utility (as measured by BLEU and ROUGE-L scores) while providing rigorous LDP guarantees, effectively preventing the extraction of private entities. This entity-level perturbation approach offers a more targeted and utility-preserving alternative to document-level DP, making it particularly suitable for contexts like healthcare, where preserving clinical semantics is critical.

\subsubsection{Synthetic Data Generation}
Synthetic data generation represents an innovative approach to privacy preservation in RAG systems by fundamentally eliminating privacy risks through the replacement of sensitive real-world data with artificially generated alternatives. This methodology offers the theoretical advantage of complete privacy protection, as no actual patient information is utilized in the retrieval or generation processes.

A work conducted by Zeng et al. \cite{zeng2025mitigating} establishes the foundational framework for implementing synthetic data approaches in RAG systems. They use generative models to create synthetic documents that replicate the structural and semantic characteristics of original datasets while introducing sufficient randomization to prevent re-identification risks. The synthetic data generation process involves careful calibration of generation parameters to maintain the relevance and accuracy of information retrieval while ensuring that no real-world sensitive information can be inferred from the synthetic alternatives. The framework incorporates utility preservation mechanisms that evaluate the quality of generated synthetic data through comprehensive similarity metrics and downstream task performance assessments. Through addressing the fundamental challenge of preserving data utility and ensuring complete privacy protection through novel synthetic data generation techniques, the researchers capture the underlying patterns and distributions of real datasets without retaining any identifiable information.

\subsubsection{Data Encryption and Anonymization}
This approach prioritizes data protection by transforming it before it enters the RAG system. Data encryption converts information into a secure ciphertext, while data anonymization replaces sensitive elements. In the event of a data breach, these measures prevent hackers from recovering the original, sensitive information.

Cryptographic methods provide the strongest, mathematically rigorous privacy guarantees by performing computations directly on encrypted data. This paradigm ensures that sensitive information remains encrypted end-to-end, securing it against exposure even to the cloud service provider. A key innovation to overcome prohibitive computational overhead is the knowledge distillation of explicit knowledge into compact, structured parametric representations, which are then used for encrypted reasoning. Chen et al. \cite{chen2025privacy} introduce DistilledPRAG, a privacy-preserving framework that enables LLMs to reason over private documents without exposing them in plaintext. The system trains a parameter generator that converts documents into corresponding Low-Rank Adaptation (LoRA) weights. These LoRA weights are then injected into the model to guide reasoning. During training, DistilledPRAG aligns its internal behavior with a standard RAG teacher model by applying knowledge distillation to both the hidden states and output logits, ensuring that the model learns to reproduce RAG-style reasoning without accessing the original text.

Similarly, Ning et al. \cite{ning2025privacy} address Privacy-protected Knowledge Graph Question Answering (KGQA) with the ARoG framework. It anonymizes all KG entities with meaningless Machine Identifiers (MIDs) and uses two abstraction strategies: Relation-centric Abstraction infers high-level concepts for anonymous entities, while Structure-oriented Abstraction converts natural language questions into structured abstract concept paths.

Lightweight encryption offers a practical alternative by using semantic transformation to create non-sensitive representations processable by standard models, achieving a balance between security and computational efficiency. He et al. \cite{he2025press} propose PRESS, a method that defends privacy by shifting the semantic embedding space. By fine-tuning the embedding model using a reinforcement learning framework, PRESS ensures that queries designed to extract private information are mapped closer to "safe," non-sensitive content. This causes the retriever to fetch irrelevant contexts for adversarial queries, effectively creating a cryptographically-inspired "one-way function" in the semantic space that is easy to compute but hard to reverse for privacy extraction.

Fang et al. \cite{fang2025guardian} introduce Guardian Angel, which employs symbolic substitution and anonymization. This framework replaces sensitive entities with patterned strings or Universally Unique Identifiers (UUIDs) before cloud processing, using In-Context Learning (ICL) to teach the LLM to reason with these symbols. The original data is stored in a local secure database to "decrypt" the final response. This method mimics a symmetric encryption/decryption pipeline, leveraging the cloud LLM for computation without ever exposing plaintext sensitive data.

Collectively, these frameworks demonstrate a sophisticated spectrum of cryptographic and crypto-inspired methods for RAG. From pure protocols that encrypt compact knowledge representations to practical proxies that use semantic shifting and token substitution, they offer a range of options for deploying secure, privacy-preserving AI in sensitive domains like healthcare.

\subsection{Strengths and limitations of applying to healthcare}
The emerging solutions for privacy-preserving RAG systems hold significant promise for advancing healthcare AI but also present notable challenges, particularly in the context of their applicability to healthcare.

A key strength of these approaches lies in their ability to offer enhanced privacy guarantees. Methods such as differential privacy and HE provide strong theoretical protections for sensitive data. These techniques ensure that patient information remains secure throughout the retrieval and generation processes, addressing pressing concerns about data confidentiality in healthcare. Cryptographic frameworks, including SMC, further reinforce privacy by enabling computations on encrypted data, thus mitigating risks of data exposure even in collaborative or distributed environments.

Another important advantage of these solutions is their potential for scalability and collaboration, particularly through FL approaches. By distributing computation across multiple institutions while keeping data local, federated systems enable multi-institutional collaboration without the need for centralized data sharing. This is particularly beneficial in healthcare, where data is often fragmented across institutions and subject to strict privacy regulations. Federated systems align well with legal and ethical requirements such as GDPR and HIPAA, while also fostering the development of robust and generalizable AI systems.

Synthetic data generation provides an additional avenue for reducing privacy risks. By replacing sensitive real-world data with artificial datasets, this approach theoretically eliminates the possibility of privacy breaches. Synthetic data can also facilitate secure model training and testing, offering a practical way to develop RAG systems without compromising patient confidentiality. Furthermore, emerging techniques like privacy-aware decoding and domain-specific knowledge distillation demonstrate adaptability to the complex requirements of healthcare, such as handling multimodal data and preserving the semantic richness of medical information.

Despite these strengths, significant limitations hinder direct healthcare application. A key issue is limited validation in clinical settings: methods proven on general datasets may not handle the complexity and safety demands of medical workflows. Consider DP in healthcare RAG. To achieve formal privacy guarantees, DP injects noise into training signals or inference-time intermediates (e.g., retrieval scores or risk estimates). Even small perturbations can cross clinical thresholds. For instance, if a biopsy recommendation is triggered at a 5\% malignancy-risk cutoff, DP noise could shift an estimate from 4.8\% to 5.3\%, changing the output from “strongly recommend against biopsy in this case” to “suggest considering biopsy.” This tiny numeric change, negligible under textual similarity, alters clinical meaning and could cause harm. Current DP mechanisms are not tuned to preserve such threshold-sensitive, clinically critical nuances under strict privacy budgets.

Another challenge is the computational overhead associated with methods such as HE and secure multi-party computation. These techniques, while providing strong privacy guarantees, are computationally intensive and may not scale efficiently in real-world healthcare environments. Resource-constrained settings, such as small clinics or rural healthcare facilities, may find it difficult to adopt these solutions due to their high infrastructure demands. Federated systems, while promising, also face issues related to communication overhead and model convergence, particularly when dealing with heterogeneous data distributed across multiple institutions.

Additionally, adapting these emerging solutions to the unique characteristics of healthcare data presents significant obstacles. Medical data is highly diverse, encompassing unstructured clinical notes, structured databases, and imaging data. Synthetic data generation, while promising, should ensure that the artificial datasets it produces accurately capture the nuances of real medical data to remain clinically useful. Similarly, federated and cryptographic systems require extensive customization to accommodate the variability in healthcare data formats and standards.






\section{Discussion}
Firstly, based on the previous sections, while current privacy-preserving approaches like on-device processing and basic FL address data exposure, they face fundamental challenges in scalability, data heterogeneity, and the privacy-utility trade-off. Emerging solutions directly target these limitations by proposing federated frameworks for scalability, synthetic data for data harmonization, and advanced cryptography for algorithmic security. However, these promising solutions inherit and even amplify the core challenges of their predecessors, particularly the computational burden of cryptographic methods and the persistent difficulty in balancing robust privacy with clinical utility. This underscores a critical need for future work to transit from theoretical privacy guarantees to practical, efficient, and clinically validated frameworks that are specifically calibrated for the high-stakes healthcare environment.

A cross-cutting theme that permeates both current and emerging solutions is the unavoidable privacy-utility trade-off, which manifests uniquely across different technical paradigms. In differential privacy, utility is traded for provable security through noise injection, which risks blurring critical clinical distinctions. Synthetic data approaches exchange the fidelity of real-world data for the theoretical safety of artificial datasets, potentially losing rare but crucial edge cases. For example, a synthetic dataset used to train a disease detection system might overlook uncommon symptoms of a serious condition. This could cause the system to miss diagnoses for some patients, leading to delayed or inappropriate treatment. FL sacrifices the convergence speed and stability of centralized training for data locality, incurring significant communication overhead and grappling with non-IID data distributions across hospitals. Finally, cryptographic methods provide the strongest security guarantees but trade off computational efficiency, making real-time, point-of-care interactions challenging. The central challenge for future healthcare RAG systems is not to eliminate this trade-off, but to develop adaptive frameworks that dynamically optimize the balance based on the sensitivity of the data being processed and the criticality of the clinical task at hand.

A critical limitation across all reviewed research is the pronounced absence of standardized objective evaluation frameworks for data privacy protection in RAG systems. Currently, the field lacks both standardized benchmarks and adaptable assessment tools, creating a significant gap in our ability to quantify privacy risks and evaluate RAG-based healthcare systems comprehensively.

Among existing privacy-preserving solutions in healthcare literature, only three incorporate data privacy evaluation. Garcia et al. \cite{garcia2025df} employ a 1-5 scale for manual expert assessment of Privacy/Security and Regulatory Compliance in system responses. However, this approach lacks objectivity, as results depend solely on expert judgment of generated answers. Jiang et al. \cite{jiang2024medirag} define "Retrieval Security" as a document-level metric to evaluate responses when implementing user access control mechanisms, assessing whether unauthorized users are prevented from accessing restricted data. This metric, while useful, is narrowly tailored to access control scenarios. Regarding edge processing, Weerasekara et al. \cite{weerasekara2025privacy} utilize only accuracy metrics to evaluate PHI sequence identification during data anonymization—a component-level assessment that does not capture whole-system privacy protection.

Similarly, emerging privacy-preserving solutions in RAG system, despite their focus on enhanced data privacy protection methods, demonstrate limited evaluation rigor. Qian et al. \cite{qian2025hyfedrag} employ GPT-4o to automatically assign 1-5 privacy scores based on whether raw personally identifiable information (PII) remains in responses. Wang et al. \cite{wang2025privacy} and Zeng et al. \cite{zeng2025mitigating} utilize metrics including Repeat Prompts and Repeat Contexts to quantify direct verbatim leakage of sensitive text, alongside ROUGE Prompts and ROUGE Contexts to assess semantic-level information disclosure, offering more granular privacy leakage analysis beyond binary outcomes. Nevertheless, these approaches remain limited: human evaluation lacks consistency and scalability, while metrics like Repeat Prompts provide only binary or sentence-level assessments. A fundamental question remains unaddressed: how can we comprehensively capture data leakage patterns across an entire response, moving beyond sentence-level granularity to identify subtle, context-dependent privacy violations? Subtle, context-dependent privacy violations matter because private information can be inferred from context rather than directly stated. The main bottleneck is that current metrics focus on explicit text overlap and fail to detect these implicit, inference-based leaks.

In addition to the lack of standardized objective evaluation metrics, there currently exists no automated assessment tool designed to evaluate privacy risks against developers' specific RAG databases and models, which is a far more complex challenge given the heterogeneity of data formats and system architectures. Since different healthcare chatbots operate on distinct datasets with varying characteristics, applying predefined benchmarks to evaluate real-world applications may prove impractical. While standardized benchmarks serve valuable purposes during the method development and validation stages, practical evaluation of deployed systems necessitates assessment on application-specific datasets to yield meaningful, context-relevant results. Furthermore, the diversity of RAG frameworks and architectural implementations across different applications introduces additional evaluation challenges, as privacy assessment tools should accommodate varying system designs, data pipeline configurations, and integration patterns.

This evaluative gap directly exacerbates regulatory and ethical challenges. Regulations like HIPAA and GDPR, focused on data access and breach notification, provide little guidance for ensuring algorithmic privacy, creating a chasm between legal compliance and technical implementation. To bridge this gap, emerging governance framework initiatives, such as the Good Machine Learning Practice (GMLP) principles published by FDA and ISO/IEC 23894 guidance on AI risk management, offer structured approaches that healthcare RAG developers can align with to ensure consistency and accountability \cite{CenterforDevicesandRadiologicalHealth, ISOIEC23894_2023}. Without standardized, quantifiable privacy metrics, healthcare organizations struggle to demonstrate compliance to regulators and institutional review boards. Furthermore, the inability to provide transparent, verifiable privacy guarantees erodes the trust of clinicians and patients, ultimately stifling the adoption of otherwise transformative technologies.

\section{Future Directions}
To address these multifaceted challenges, future work should extend beyond algorithmic innovation to establish a robust, practical, and tiered evaluation ecosystem that directly bridges the gap between policy and technology. We propose the following critical directions as a foundational checklist for future research and development:
\begin{enumerate}
    \item \textbf{Develop Automated Privacy Assessment Tools for Chatbots} There is an urgent need to create flexible, automated tools specifically designed to evaluate the data privacy protection capability of a deployed chatbot. These tools should be adaptable to diverse system architectures and should automatically simulate real-world privacy attacks, providing a quantifiable privacy risk score for any given RAG pipeline.

    \item \textbf{Establish Standardized Metrics and Definitions for Data Leakage} The field should move towards a consensus on what constitutes a "data leakage" in the context of RAG. This involves creating clear, measurable definitions and standardized metrics—such as resilience against membership inference, prompt extraction, or training data reconstruction attacks—to replace the current qualitative and often vague assessments.

    \item \textbf{Implement a Tiered Data Sensitivity Framework} A one-size-fits-all approach to privacy is inefficient. Future frameworks should define a multi-level taxonomy for classifying healthcare data (e.g., public, de-identified, limited dataset, fully identified PHI). Privacy preservation efforts and computational resources can then be allocated proportionally to the sensitivity level, ensuring robust protection for critical information without unnecessary overhead.

    \item \textbf{Bridge the Policy-Technical Divide} Crucially, the above technical tools should be explicitly linked to regulatory compliance. This involves translating legal principles from HIPAA and GDPR into concrete, measurable technical specifications. For instance, a "right to be forgotten" should be operationalized into a verifiable data deletion protocol within a RAG system's architecture. The goal is to create a certification framework where passing automated privacy assessments and adhering to the tiered data framework would serve as auditable evidence of compliance, moving from abstract legal obligations to demonstrable technical safeguards.
\end{enumerate}

The path forward requires a concerted effort to translate these proposed directions from a theoretical checklist into a practical, industry-wide standard. Continued collaboration among researchers, developers, and regulators will be essential in shaping an integrated evaluation ecosystem. Embedding consistent, multi-layered privacy assessments throughout the development process can help ensure that healthcare RAG systems operate reliably and earn patient trust.

\section{Conclusions}
This review has synthesized the landscape of Retrieval-Augmented Generation in healthcare, revealing a critical tension between its potential for clinical applications and the privacy risks inherent in its architecture. RAG-LLMs have the potential to transform healthcare by delivering accurate, evidence-based insights. However, realizing this potential depends on effectively ensuring data privacy. This review therefore serves as a key contribution, outlining the vulnerabilities and safeguards that will shape the future direction of the field. By systematically analyzing vulnerabilities across the RAG pipeline and critically evaluating both current and emerging privacy-preserving solutions, a central theme emerges: technical efficacy and robust data protection are inseparable requirements. While solutions from data localization to advanced cryptography offer promising pathways, they are hampered by the privacy-utility trade-off, computational overhead, and a critical lack of standardized evaluation. Therefore, the future of trustworthy healthcare RAG lies not in any single technical silver bullet, but in a holistic approach that integrates adaptive, context-aware privacy frameworks with rigorous, standardized assessments. Ultimately, this work underscores that prioritizing patient privacy is not a constraint on innovation but the foundation upon which clinically effective, ethically sound, and widely adopted AI-driven healthcare should be built.

\bibliographystyle{unsrt}  
\bibliography{custom.bib}  

\end{document}